\documentclass[preprint]{revtex4}
\usepackage{epsfig}

\def\beq{\begin{equation}}
\def\eeq{\end{equation}}
\def\beqa{\begin{eqnarray}}
\def\eeqa{\end{eqnarray}}

\begin{document}

\title{\sc Non-linear waves in a Quark Gluon Plasma}

\author{D.A. Foga\c{c}a\dag\, L.G. Ferreira Filho\ddag\   and F.S. Navarra\dag\ }
\address{\dag\ Instituto de F\'{\i}sica, Universidade de S\~{a}o Paulo\\
 C.P. 66318,  05315-970 S\~{a}o Paulo, SP, Brazil}
\address{\ddag\ Faculdade de Tecnologia, Universidade do Estado do Rio de Janeiro \\
Via Dutra km 298, CEP 27523-000, Resende, RJ, Brazil}

\begin{abstract}

Recent measurements at RHIC suggest that  a nearly perfect fluid of quarks and 
gluons is produced in $A \,A$ collisions. Moreover  the passage of supersonic partons 
through this medium seems to produce waves. These waves might pile up and 
form Mach cones, which would manifest themselves in the so called away-side jets, forming 
a broad structure in the angular distribution of the particles recoiling against a trigger  
jet of moderate energy. In most of the theoretical descriptions of these phenomena, 
the hydrodynamic equations are linearized for simplicity. We propose an alternative explanation 
for the observed broadening of the away-side peak. It is based on hydrodynamics but it is a 
consequence of the non-linearities of the equations, which instead of simple waves may lead to 
localized waves or even solitons. 

We investigate in detail the 
consequences of including the non-linear terms. We use a simple equation of 
state for the QGP and expand the hydrodynamic equations around equilibrium 
configurations. The resulting differential equations describe the propagation of 
perturbations in the energy density. We solve them numerically and find that 
localized perturbations can propagate for long distances in the plasma. Under 
certain conditions our solutions mimick the propagation of  Korteweg - de Vries 
solitons.
\end{abstract} 

\maketitle



\vspace{1cm}
\section{Introduction}

The heavy ion collisions performed at BNL's  Relativistic heavy Ion Collider (RHIC) 
create a hot and dense  medium, which behaves as a perfect fluid. During the first 
years of the RHIC program, hydrodynamics was applied to describe the space-time evolution 
of the bulk of the fluid. In the last years  hydrodynamics became relevant  
to study also the perturbations on the fluid, such as, for example, the  waves generated by 
the passage of a supersonic parton.  This field was opened by the observation of a 
broad structure  in azimuthal di-hadron correlations \cite{star,phenix}. This 
broad  structure is called the  ``away-side jet'' and  recoils against the ``near-side jet''
(or ``trigger jet''). In the framework of hydrodynamics, this observation could be explained by 
the conical shock waves generated by large energy deposition in the hydrodynamical medium
\cite{ma1,ma2,ma3,ma4,ma5,ma6,ma7,ma8,ma9,betz1,betz2,betz3}.  Although quite elegant, this 
understanding of the away-side jet in terms of conical shock waves still needs confirmation.
A very recent and improved analysis by the STAR collaboration has given further support to 
this picture \cite{star09}. A more solid evidence of this phenomenon   may come from the 
study of 
jets at the Large Hadron Colider (LHC), where the energy released by the nuclear projectiles 
in the central rapidity region  will be larger \cite{dnw} and so the formed fireball will be 
larger and live longer, allowing for a more complete  study of waves. 

In this work we discuss another possible mechanism for the formation of broad structures in 
the away-side jet. In the limit where the jet looses most of its energy, which is rapidly 
thermalized and incorporated to fluid, a pulse is formed, which propagates through the fluid. 
During its motion this energy density pulse spreads both in the longitudinal and transverse 
direction. After hadronization this travelling and expanding ``hot spot'' will form particles 
with a broader angular distribution than those coming from the near-side jet. This is depicted 
in Fig. 1. Notice that in this process there is no Mach cone formation.  During the motion of 
the energy density pulse, the medium undergoes an expansion leading to a spread of this pulse. 
A further spreading will occur during the hadronization and final particle formation.  Therefore, 
in this picture it is essential that the initial perturbation remains localized to a good extent. 
Otherwise it will spread too much and destroy the jet-like topology, which is compatible with 
data.  
\begin{figure}[h]
\begin{center}
\epsfig{file=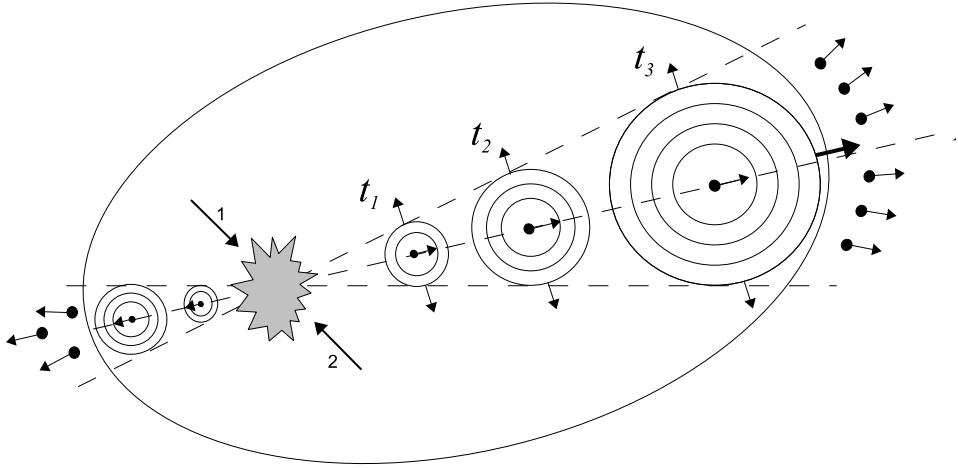,height=100mm}
\caption{Parton-parton collision forming two back-to-back jets, which evolve in a hot quark gluon 
plasma. The circles represent  localized (soliton-like) energy density perturbations which 
traverse the fluid and suffer expansion, forming a narrow near-side jet and a broad 
away-side jet.} 
\label{fig1}
\end{center}
\end{figure}
Highly localized perturbations can exist and  propagate through a fluid. The most famous are the 
Korteweg - de Vries (KdV) solitons, which are solutions of the KdV equation. This equation  may 
be derived from the equations of hydrodynamics under certain conditions. One of them is to 
preserve the non-linear terms of the Euler and continuity equations. The other one is to have a 
third order spatial derivative term. This term comes from the equation of state of the fluid 
and it appears because the Lagrangian density contains higher derivative couplings
\cite{fn1,fn2,fn3} or because of 
the Laplacians appearing in the equations of motion of the fields of the theory \cite{fn4}.  
This happens, 
for example, in the non-linear Walecka model of nuclear matter at zero and finite temperature. 
For a quark gluon plasma (QGP) it depends on the  coupling regime and on the properties of the 
QCD vacuum. As it will be seen in this work, if we consider the simplest case of a free gas of 
massless quarks and gluons, the hydrodynamical equations do not give origin to the KdV equation.  
Instead they generate a non-linear differential equation for the  perturbation which has no 
third order stabilising term. This equation is called wave breaking equation and is also very 
well knonw in the literature. The numerical solution of this equation shows that an initial
gaussian-like perturbation in the energy density  evolves creating a vertical ``wall'' in its 
front, which breaks and looses localization.  In our case, surprisingly enough, this same 
phenomenon happens but it takes a very long time and long distances, compared to the nuclear 
scales. So, from the practical point of view, there is no distinction between a breaking pulse
and a soliton. This persistence of localization in the breaking wave is the main result of our
paper and  gives support to the process shown in Fig. 1. 
However, from this finding to a 
realistic calculation and a serious  attempt to describe the data there is still a long way. 
The next step now will be to quantify the broadening of the moving bubble in Fig. 1, which will 
be directly reflected in the angular distribution of the fragments.  For this we need to extend 
our formalism 
to two spatial dimensions (longitudinal $x$ and radial $r$). This is a  heavily numerical
project and it is still in progress. Based on previous works with the analogous 
non-relativistic  problem for nuclear matter, discussed in \cite{rww}, we have reasons to expect 
a soliton-like evolution along the $x$ direction with a ``leakage'' to the radial direction, 
which  would cause the angular  broadening in the final matter distribution.

In the  theoretical description of  these perturbations \cite{ma2,ma8,ma9}, 
very often the  hydrodynamic equations are linearized for simplicity. 
As it is usually done in non-relativistic hydrodynamics, 
linearization consists \cite{hidro1} in  considering only first order terms in the 
velocity and in the  energy  and pressure perturbations and neglecting higher order 
terms and derivatives involving them. 
In this work we revisit the relativistic hydrodynamic equations 
expanding them in a different way, in terms of a small expansion parameter ($\sigma$)  closely 
following what
is done in magnetohydrodynamics of plasmas \cite{davidson} and keeping the non-linear 
features of the problem. Techniques of plasma physics started to be  applied to  nuclear 
hydrodynamics  long ago  \cite{frsw,abu} to study perturbations in the cold nucleus, 
treated as a fluid. 
We extended those pioneering studies to relativistic and warm nuclear matter
\cite{fn1,fn2,fn3,fn4} and now to the quark gluon plasma (QGP). 

The most interesting aspect of  \cite{frsw,abu,fn1,fn2,fn3,fn4} was to find at 
some point of the developement, the (KdV) equation for the perturbation in the  
nuclear  matter density.  This is  the ``nuclear soliton''. Our main contribution 
was to establish a connection between the KdV equation (and the properties of its 
solitonic solutions) and a modern underlying nuclear matter theory (which in our  
case was a variant of  the non-linear Walecka model)  and then to show 
that the soliton solution exists even in relativistic hydrodynamics \cite{fn1,fn2}.

In the next section we review the main formulas of relativistic hydrodynamics. In 
section III we discuss the quark gluon plasma equation of state. In sections IV and 
V we show how to derive the diffferential equations which govern the time evolution of 
perturbations at zero and finite temperature respectively.  In section VI we present 
the numerical solutions of the obtained differential equations and in section VII we present
some conclusions.

\section{Relativistic Fluid Dynamics}

In this section we review the main expressions of one dimensional relativistic 
hydrodynamics. Throughout this work we employ  natural units $c=1$, $\hbar=1$ and  
(Boltzmann's constant) $k_{B}=1$. The velocity four vector  $u^{\nu}$ is defined as 
$u^{0}=\gamma$,  $\vec{u}=\gamma \vec{v}$, 
where $\gamma$ is the Lorentz factor given by 
$\gamma=(1-v^{2})^{-1/2}$ and thus $u^{\nu}u_{\nu}=1$.  The velocity field of the 
matter is $\vec{v}=\vec{v}(t,x,y,z)$.  The energy-momentum tensor is, as usual, 
given by:
\begin{equation}
T_{\mu \nu}=(\varepsilon +p)u_{\mu}u_{\nu}-pg_{\mu\nu}
\label{tensor}
\end{equation}
where  $\varepsilon$ and $p$ are the energy density and pressure respectively. 
Energy-momentum conservation is  ensured by:
\begin{equation}
\partial_{\nu}{T_{\mu}}^{\nu}=0
\label{cons}
\end{equation}
The projection of (\ref{cons}) onto a direction perpendicular to $u^{\mu}$ gives 
the relativistic version of  the Euler equation \cite{wein,land}:
\begin{equation}
{\frac{\partial {\vec{v}}}{\partial t}}+(\vec{v} \cdot \vec{\nabla})\vec{v}=
-{\frac{1}{(\varepsilon + p)\gamma^{2}}}
\bigg({\vec{\nabla} p +\vec{v} {\frac{\partial p}{\partial t}}}\bigg)
\label{eul}
\end{equation}
The relativistic version of the continuity equation for the baryon density is 
\cite{wein}:
\begin{equation}
\partial_{\nu}{j_{B}}^{\nu}=0
\label{conucleon}
\end{equation}
Since ${j_{B}}^{\nu}=u^{\nu} \rho_{B}$ the above equation can be rewritten as:
\begin{equation}
{\frac{\partial \rho_{B}}{\partial t}}+\gamma^{2}v \rho_{B}\Bigg({\frac{\partial v}
{\partial t}}+ \vec{v}\cdot \vec{\nabla} v\Bigg)+\vec{\nabla} \cdot (\rho_{B}\vec{v})=0
\label{rhobcons}
\end{equation}
The relativistic version of the continuity equation for the entropy density is given
by the projection of (\ref{cons}) onto  the direction of $u^{\nu}$ \cite{land}:
\begin{equation}
(\varepsilon+p)\partial_{\mu}u^{\mu}+u^{\mu}\partial_{\mu}\varepsilon=0
\label{gconspro}
\end{equation} 
At this point we recall the Gibbs relation:
\begin{equation}
\varepsilon + p={\mu_{B}}{\rho_{B}}+Ts
\label{gibaoporV}
\end{equation} 
and the first law of thermodynamics:
\begin{equation}
d\varepsilon=Tds+{\mu_{B}}d{\rho_{B}}
\label{primlei}
\end{equation}
We will later consider a hot gas of quarks and gluons, where the net baryon density 
is zero, i.e.,   
$\rho_{B}=0$ ($d\rho_{B}=0$) at $T\neq 0$. Using this last relation in (\ref{primlei})  
and then inserting (\ref{primlei}) and (\ref{gibaoporV}) in (\ref{gconspro}) we arrive 
at 
$$
Ts(\partial_{\mu}u^{\mu})+Tu^{\mu}(\partial_{\mu}s)=0
$$
and finally at
\begin{equation}
\partial_{\nu}(s{u}^{\nu})=0
\label{scons}
\end{equation}
which was expected for a perfect fluid. For future use, the above formula will be 
expanded as:
\begin{equation}
{\frac{\partial s}{\partial t}}+\gamma^{2}v s\Bigg({\frac{\partial v}
{\partial t}}+ \vec{v}\cdot \vec{\nabla} v\Bigg)+\vec{\nabla} \cdot (s\vec{v})=0
\label{sconsc}
\end{equation}
which is quite similar to (\ref{rhobcons}).

\section{ The  QGP Equation of State} 

We shall use a simple equation of state derived from the MIT Bag Model. It describes 
an ideal gas of quarks and gluons and takes into account the effects of confinement 
through the bag constant $\mathcal{B}$. This constant is interpreted as the energy 
needed to create a bubble or bag in the vacuum (in which the noninteracting quarks 
and gluons are confined) and it can  be extracted from hadron spectroscopy  or from 
lattice QCD calculations. There is a relationship between  
$\mathcal{B}$ and the critical temperature of the quark-hadron transition $T_{c}$ 
which is determined by assuming that, during the  phase transition, the pressure 
vanishes.   

The baryon density is given by: 
\begin{equation}
\rho_{B}={\frac{1}{3}}{\frac{\gamma_Q}{(2\pi)^{3}}}\int d^3{k}\hspace{0.2cm}
[n_{\vec{k}}-\bar{n}_{\vec{k}}]
\label{rodensTq}
\end{equation}
where
\begin{equation}
n_{\vec{k}} \equiv n_{\vec{k}}(T)={\frac{1}{1+e^{(k-{\frac{1}{3}}\mu)/ T}}}
\label{qdis}
\end{equation}
and
\begin{equation}
\bar{n}_{\vec{k}} \equiv \bar{n}_{\vec{k}}(T)={\frac{1}{1+e^{(k+{\frac{1}{3}}\mu)/ T}}}
\label{aqdis}
\end{equation}
where from now on $\mu$ is the baryon chemical potential. At zero temperature the expression for 
the baryon density reduces to:
\begin{equation}
{\rho_{B}}={\frac{2}{3\pi^{2}}}{k_{F}}^{3} 
\label{a12pa}
\end{equation}
where  $k_{F}$ is the highest occupied  level. 
The energy density and the pressure are given by:
\begin{equation}
\varepsilon=\mathcal{B}+{\frac{\gamma_G}{(2\pi)^{3}}}\int d^3{k}\hspace{0.2cm}k\hspace{0.2cm}(e^{k/T}-1)^{-1}
+{\frac{\gamma_Q}{(2\pi)^{3}}}\int d^3{k}\hspace{0.2cm}k\hspace{0.2cm}[n_{\vec{k}}
+\bar{n}_{\vec{k}}]
\label{esdTqg}
\end{equation}
and
\begin{equation}
p=-\mathcal{B}+{\frac{1}{3}}\Bigg\lbrace {\frac{\gamma_G}{(2\pi)^{3}}}\int d^3{k}\hspace{0.2cm}k\hspace{0.2cm}(e^{k/T}-1)^{-1}+
{\frac{\gamma_Q}{(2\pi)^{3}}}\int d^{3}k\hspace{0.1cm}k\bigg[{n}_{\vec{k}}+\bar{n}_{\vec{k}}\bigg] \Bigg\rbrace
\label{psdTqg}
\end{equation}
The statistical factors  are 
$\gamma_G=2\textrm{(polarizations)}\times 8\textrm{(colors)}=16$ for gluons and 
$\gamma_Q=2\textrm{(spins)}\times 2\textrm{(flavors)}\times 3
\textrm{(colors)}=12$ for quarks. From the above expressions we derive the useful 
formulas:
\begin{equation}
3(p+\mathcal{B})=\varepsilon-\mathcal{B}={\frac{8\pi^{2}}{15}} 
\ T^{4}+{\frac{6}{\pi^{2}}}\int_{0}^{\infty} d{k}\hspace{0.2cm}k^{3}
[n_{\vec{k}}+\bar{n}_{\vec{k}}]
\label{bacana}
\end{equation}
and
\begin{equation}
p={\frac{1}{3}}\varepsilon-{\frac{4}{3}}\mathcal{B}
\label{eosqg}
\end{equation}
The  speed of sound, $c_{S}$, is given by
\begin{equation}
{c_{S}}^{2}={\frac{\partial p}{\partial \varepsilon}}={\frac{1}{3}}
\label{soundone}
\end{equation}

\section{Wave equation at zero temperature}

In the core of a dense star the temperature is close to zero and the baryon density is very 
high. The quark distribution function becomes the step function.
Using (\ref{a12pa}) in  (\ref{esdTqg}) and (\ref{psdTqg}) we find:
\begin{equation}
\varepsilon(\rho_{B})=\bigg({\frac{3}{2}}\bigg)^{7/3}\pi^{2/3}{\rho_{B}}^{4/3}
+\mathcal{B}
\label{bacanatrh}
\end{equation}
and
\begin{equation}
p(\rho_{B})={\frac{1}{3}}\bigg({\frac{3}{2}}\bigg)^{7/3}\pi^{2/3}
{\rho_{B}}^{4/3}-\mathcal{B}
\label{prerho}
\end{equation}
From (\ref{eosqg}) we have 
$\vec{\nabla}p={\frac{1}{3}}\vec{\nabla}\varepsilon$ and also
$ {\frac{\partial p}{\partial t}} = 
{\frac{1}{3}}{\frac{\partial \varepsilon}{\partial t}}$. Combining these expressions
with (\ref{bacanatrh}) and  (\ref{prerho}) we find:
\begin{equation}
\vec{\nabla}p={\frac{4}{9}}\bigg({\frac{3}{2}}\bigg)^{7/3}\pi^{2/3}{\rho_{B}}^{1/3} 
\ \vec{\nabla}{\rho_{B}}
\label{pevar}
\end{equation}
and
\begin{equation}
{\frac{\partial p}{\partial t}}={\frac{4}{9}}\bigg({\frac{3}{2}}\bigg)^{7/3}\pi^{2/3}{\rho_{B}}^{1/3} \ 
{\frac{\partial {\rho_{B}}}{\partial t}}
\label{ptvar}
\end{equation}
Finally, substituting (\ref{bacanatrh}), (\ref{prerho}),  (\ref{pevar}) and 
(\ref{ptvar}) into  (\ref{eul}) we obtain:
\begin{equation}
{\rho_{B}}\bigg[{\frac{\partial {\vec{v}}}{\partial t}}+(\vec{v} \cdot 
\vec{\nabla})\vec{v}\bigg]=
\frac{(v^{2}-1)}{3}\bigg[\vec{\nabla}{\rho_{B}}+\vec{v}{\frac{\partial 
{\rho_{B}}}{\partial t}}\bigg]
\label{eulerqhdqcd}
\end{equation}
which is the relativistic version of Euler equation for the QGP at $T=0$.

Following the same formalism already used for nuclear matter in 
\cite{fn1,fn2,fn3,fn4} we will now expand both (\ref{rhobcons}) and 
(\ref{eulerqhdqcd}) in powers of a small parameter $\sigma$ and combine these two 
equations to find one single differential equation which governs the space-time 
evolution of the perturbation in the baryon density. 
We write (\ref{rhobcons})  and (\ref{eulerqhdqcd}) in one cartesian dimension 
($x$) in terms of the dimensionless variables:
\begin{equation}
\hat\rho={\frac{\rho_{B}}{\rho_{0}}} \hspace{0.2cm}, \hspace{0.5cm} \hat v={\frac{v}{c_{S}}}
\label{vadima}
\end{equation}
where $\rho_0$ is an equilibrium (or reference) density, upon which perturbations may be 
generated. Next, we introduce the  $\xi$ and $\tau$ ``stretched'' coordinates 
\cite{frsw,abu,davidson}:
\begin{equation}
\xi=\sigma^{1/2}{\frac{(x-{c_{S}}t)}{R}} 
\hspace{0.2cm}, \hspace{0.5cm} 
\tau=\sigma^{3/2}{\frac{{c_{S}}t}{R}} 
\label{streta}       
\end{equation} 
After this change of variables  we  expand   (\ref{vadima}) as:
\begin{equation}
\hat\rho=1+\sigma \rho_{1}+ \sigma^{2} \rho_{2}+ \dots
\label{roexpa}
\end{equation}
\begin{equation}
\hat v=\sigma v_{1}+ \sigma^{2} v_{2}+ \dots
\label{vexpa}
\end{equation} 
Neglecting terms proportional to $\sigma^{n}$ for $n\geq 3$ and organizing the 
equations as series in powers of $\sigma$,
(\ref{rhobcons}) and (\ref{eulerqhdqcd}) aquire the form:
$$
\sigma \Bigg\lbrace {\frac{\partial\rho_{1}}{\partial \xi}}-{\frac{\partial v_{1}}{\partial \xi}} \Bigg\rbrace+
\sigma^{2} \Bigg\lbrace {\frac{\partial v_{2}}{\partial \xi}}-{\frac{\partial\rho_{2}}{\partial \xi}}
+{\frac{\partial\rho_{1}}{\partial \tau}}+\rho_{1}{\frac{\partial v_{1}}{\partial \xi}}+v_{1}{\frac{\partial \rho_{1}}{\partial \xi}}
-{c_{S}}^{2}v_{1}{\frac{\partial v_{1}}{\partial \xi}} \Bigg\rbrace =0 
$$
and
$$
\sigma \Bigg\lbrace {\frac{1}{3{c_{S}}^{2}}}{\frac{\partial\rho_{1}}{\partial \xi}}-{\frac{\partial v_{1}}{\partial \xi}} \Bigg\rbrace+
\sigma^{2} \Bigg\lbrace -{\frac{\partial v_{2}}{\partial \xi}}+{\frac{1}{3{c_{S}}^{2}}}{\frac{\partial\rho_{2}}{\partial \xi}}
+{\frac{\partial v_{1}}{\partial \tau}}+v_{1}{\frac{\partial v_{1}}{\partial \xi}}-2\rho_{1}{\frac{\partial v_{1}}{\partial \xi}}
-{\frac{v_{1}}{3}}{\frac{\partial \rho_{1}}{\partial \xi}}+{\frac{\rho_{1}}{3{c_{S}}^{2}}}{\frac{\partial \rho_{1}}{\partial \xi}} \Bigg\rbrace =0 
$$
respectively. In these equations each bracket must vanish independently, i.e. 
$\lbrace \dots  \rbrace = 0$. From  the terms proportional to $\sigma$ we 
obtain ${c_{S}}^{2}=1/3$ and 
$\rho_{1}=v_{1}$, which are then inserted into the terms 
proportional to $\sigma^{2}$ giving after some algebra:
\begin{equation}
{\frac{\partial\rho_{1}}{\partial \tau}}
+{\frac{2}{3}}\rho_{1}{\frac{\partial \rho_{1}}{\partial \xi}}=0 
\label{bwqcdxitau}
\end{equation}
Returning  to the  $x-t$ space the above equation reads:
\begin{equation}
{\frac{\partial\hat\rho_{1}}{\partial t}}+ 
c_{S}{\frac{\partial \hat\rho_{1}}{\partial x}}+
{\frac{2}{3}}c_{S}\hat\rho_{1}{\frac{\partial \hat\rho_{1}}{\partial x}}=0 
\label{bwqcdxitauXt}
\end{equation}
where we have used the notation  $\hat\rho_{1}\equiv \sigma\rho_{1}$, which  is a 
small perturbation in the  baryon density.  
The equation (\ref{bwqcdxitauXt}) is the so called  breaking wave
equation for $\hat\rho_{1}$ at zero temperature in the QGP.

\section{Wave equation at finite temperature}

In the central rapidity region of a typical heavy ion collision at RHIC we have a vanishing 
net baryon number, i.e.,  $\rho_{B}=0$. The energy is mostly stored in the gluon field, which 
forms the hot and dense medium. We will now aply hydrodynamics to study this medium and focus 
on perturbations in the energy density and their propagation. Following the formalism developed 
in the previous section we will expand and combine the Euler equation given by (\ref{eul}) and 
the continuity equation for the entropy density given by (\ref{sconsc}).

As $\rho_{B}=0$, the baryon chemical potential is zero $(\mu=0)$ and so the distribution 
functions given by (\ref{qdis}) and (\ref{aqdis})
are the same, i.e. : ${n}_{\vec{k}}=\bar{n}_{\vec{k}} = {1}/{(1+e^{k/ T})}$. In this case 
the integral in (\ref{bacana}) can be easily performed and we obtain:
\begin{equation}
3(p+\mathcal{B})=\varepsilon-\mathcal{B}={\frac{37}{30}} \pi^{2} T^{4}
\label{bacanaaend}
\end{equation}
Solving the first identity for the pressure and recalling  \cite{reif} 
that $s= ({{\partial p}/{\partial T}})_{V}$ we arrive at:
\begin{equation}
s={\frac{\partial }{\partial T}}\bigg(-\mathcal{B}+{\frac{37}{90}} \pi^{2} T^{4}\bigg)=
4  \\ {\frac{37}{90}} \pi^{2} T^{3}
\label{denstemp}
\end{equation}

The ``bag constant'' parameter, $\mathcal{B}$, can be replaced by 
the critical temperature of the quark-hadron transition $T_{c}$.  
When $p=0$ at the phase transition, (\ref{bacanaaend}) reduces to:
\begin{equation}
\mathcal{B}={\frac{37}{90}} \pi^{2} (T_{c})^{4}
\label{BT}
\end{equation}
Inserting the above equation into the second identity of (\ref{bacanaaend}) we have 
the following expression for $\varepsilon(T)$:
$$
\varepsilon={\frac{37}{30}} \pi^{2}\Bigg(T^{4}+{\frac{{T_{c}}^{4}}{3}}\Bigg)
$$
From this formula we can define  the reference energy density $\varepsilon_{0}$, which  
is related to a  reference temperature, $T_{0}$, through:
\begin{equation}
\varepsilon_{0}={\frac{37}{30}} \pi^{2}\Bigg({T_{0}}^{4}+{\frac{{T_{c}}^{4}}{3}}\Bigg)
\label{densref}
\end{equation}

Solving the second identity of  (\ref{bacanaaend}) for the temperature we obtain:
\begin{equation}
T=\Bigg[{\frac{30}{37\pi^{2}}}(\varepsilon-\mathcal{B})\Bigg]^{1/4}
\label{TfromEpsilon}
\end{equation}
which, inserted  into (\ref{denstemp}) yields:
\begin{equation}
s=s(\varepsilon)=4  \\ {\frac{37}{90}} \pi^{2} \Bigg[{\frac{30}{37\pi^{2}}}
(\varepsilon-\mathcal{B})\Bigg]^{3/4}
\label{densenerd}
\end{equation}
Substituing then (\ref{densenerd}) in (\ref{sconsc}) in the one dimensional case 
and using (\ref{bacanaaend}) to write $(\varepsilon-\mathcal{B})$ in terms of the 
temperature we have finally:
\begin{equation}
(1-v^{2})\Bigg[\Bigg({\frac{90}{148\pi^{2} T^{4}}}\Bigg){\frac{\partial \varepsilon}{\partial t}}
+{\frac{\partial v}{\partial x}}+
\Bigg({\frac{90v}{148\pi^{2} T^{4}}}\Bigg){\frac{\partial \varepsilon}{\partial x}}\Bigg]+
v \Bigg({\frac{\partial v}{\partial t}}+v{\frac{\partial v}{\partial x}}\Bigg)=0
\label{sconscumdimagain}
\end{equation}
Also from (\ref{bacanaaend}) we have
\begin{equation}
\varepsilon+p={\frac{148}{90}} \pi^{2} T^{4}
\label{eps+pe}
\end{equation}
Inserting the above equation  into (\ref{eul}) and using 
$\vec{\nabla}p={\frac{1}{3}}\vec{\nabla}\varepsilon$ and also 
$ {\frac{\partial p}{\partial t}} = {\frac{1}{3}}{\frac{\partial \varepsilon}{\partial t}}$ 
we find:
\begin{equation}
{\frac{148}{30}} \pi^{2} T^{4}\bigg({\frac{\partial v}{\partial t}}+
v{\frac{\partial v}{\partial x}}\bigg)=
(v^{2}-1)\bigg({\frac{\partial \varepsilon}{\partial x}}+v{\frac{\partial \varepsilon}
{\partial t}}\bigg)
\label{eulerqgpT}
\end{equation}

We now rewrite (\ref{sconscumdimagain}) and (\ref{eulerqgpT}) in dimensionless variables:
\begin{equation}
\hat{\varepsilon}={\frac{\varepsilon}{\varepsilon_{0}}} \hspace{0.2cm}, \hspace{0.5cm} \hat v={\frac{v}{c_{S}}}
\label{vadimaft}
\end{equation}
where $\varepsilon_{0}$ is the reference  energy density given by (\ref{densref}) . 
Expanding  (\ref{vadimaft}) in powers of $\sigma$ we have:
\begin{equation}
\hat\varepsilon=1+\sigma \varepsilon_{1}+ \sigma^{2} \varepsilon_{2}+ \dots
\label{epoexpa}
\end{equation}
and
\begin{equation}
\hat v=\sigma v_{1}+ \sigma^{2} v_{2}+ \dots
\label{vexpa}
\end{equation}
Neglecting higher order terms in  $\sigma$ and changing variables to the 
$(\xi-\tau)$ space the equations  (\ref{sconscumdimagain}) and (\ref{eulerqgpT}) become:
$$
\sigma \Bigg\lbrace -{\frac{90\ \varepsilon_{0}}{148\pi^{2} T^{4}}}
{\frac{\partial\varepsilon_{1}}{\partial \xi}}+
{\frac{\partial v_{1}}{\partial \xi}} \Bigg\rbrace +
$$
\begin{equation}
\sigma^{2} \Bigg\lbrace {\frac{90 \ \varepsilon_{0}}{148\pi^{2} T^{4}}} 
\bigg(-{\frac{\partial\varepsilon_{2}}{\partial \xi}}+
{\frac{\partial\varepsilon_{1}}{\partial \tau}}+v_{1}{\frac{\partial 
\varepsilon_{1}}{\partial \xi}} \bigg)+
{\frac{\partial v_{2}}{\partial \xi}}-{c_{S}}^{2}v_{1}{\frac{\partial v_{1}}{\partial \xi}}
\Bigg\rbrace =0 
\label{entrocon}
\end{equation}
and
$$
\sigma \Bigg\lbrace -{\frac{148\pi^{2} T^{4}c_{S}}{30}}{\frac{\partial v_{1}}{\partial \xi}}+
{\frac{\varepsilon_{0}}{c_{S}}}{\frac{\partial\varepsilon_{1}}{\partial \xi}} \Bigg\rbrace  +  
$$
\begin{equation}
\sigma^{2}\Bigg\lbrace {\frac{148\pi^{2} T^{4}c_{S}}{30}} \bigg(-{\frac{\partial v_{2}}
{\partial \xi}}+{\frac{\partial v_{1}}{\partial \tau}}+
v_{1}{\frac{\partial v_{1}}{\partial \xi}}\bigg)+{\frac{\varepsilon_{0}}{c_{S}}}
{\frac{\partial \varepsilon_{2}}{\partial \xi}} -
\varepsilon_{0}{c_{S}}v_{1}{\frac{\partial \varepsilon_{1}}{\partial \xi}}  \Bigg\rbrace=0
\label{enercon}
\end{equation}
As before, in the above equations each bracket must vanish independently. 
From the first bracket of 
(\ref{entrocon}) we have:
\begin{equation}
v_{1}={\frac{90\varepsilon_{0}}{148\pi^{2}T^{4}}}\varepsilon_{1}
\label{vitfinita}
\end{equation}
which,  inserted into the terms proportional to $\sigma^{2}$, yields:
\begin{equation}
{\frac{\partial\varepsilon_{1}}{\partial \tau}}+
\Bigg({\frac{90\varepsilon_{0}}{148\pi^{2}T^{4}}}\Bigg){\frac{2}{3}}
\varepsilon_{1}{\frac{\partial \varepsilon_{1}}{\partial \xi}}=0 
\label{bwqcdxitauedp}
\end{equation}
Coming back to the $x-t$  space  the above equation becomes:
\begin{equation}
{\frac{\partial\hat\varepsilon_{1}}{\partial t}}+c_{S}{\frac{\partial 
\hat\varepsilon_{1}}{\partial x}}+
\Bigg({\frac{90\varepsilon_{0}}{148\pi^{2}T^{4}}}\Bigg)
{\frac{2}{3}}c_{S}\hat\varepsilon_{1}{\frac{\partial \hat\varepsilon_{1}}{\partial x}}=0 
\label{bwqcdTfin}
\end{equation}
where $\hat\varepsilon_{1}\equiv \sigma\varepsilon_{1}$ is a small perturbation in the 
energy density.
Equation (\ref{bwqcdTfin}) is the breaking  wave equation for $\hat\varepsilon_{1}$ in a QGP
at finite temperature. For our purposes we will rewrite this expression in a slightly 
different form.  Using   (\ref{densref}) and the relations deduced in the previous section,
(\ref{bwqcdTfin}) becomes finally:
\begin{equation}
{\frac{\partial\hat\varepsilon_{1}}{\partial t}}+c_{S}{\frac{\partial \hat\varepsilon_{1}}{\partial x}}+
\Bigg[1+{\frac{1}{3}}\Bigg({\frac{T_{c}}{T_{0}}}\Bigg)^{4}\Bigg]{\frac{c_{S}}{2}}
\hat\varepsilon_{1}{\frac{\partial \hat\varepsilon_{1}}{\partial x}}=0 
\label{bwqcdTfincomrefens}
\end{equation}
where $T_0 > T_c$.

\section{Numerical analysis and discussion}

The equations (\ref{bwqcdxitauXt}) and (\ref{bwqcdTfincomrefens}) have the form
\begin{equation}
{\frac{\partial f}{\partial t}}+c_{S}{\frac{\partial f}{\partial x}}+\alpha 
f{\frac{\partial f}{\partial x}}=0
\label{bwqcdToutzgeral}
\end{equation} 
which is a particular case of the equation:
\begin{equation}
{\frac{\partial f}{\partial t}}+c_{S}{\frac{\partial f}{\partial x}}+
\alpha f{\frac{\partial f}{\partial x}}+
B{\frac{\partial^{3} f}{\partial x^{3}}}=0
\label{bwqcdToutzgeralKDV}
\end{equation}
when $B=0$.  The last equation is the famous Korteweg - de Vries (KdV) equation, 
which has an analytical soliton solution given by \cite{drazin}:
\begin{equation}
f(x,t)={\frac{3(u-c_{S})}{\alpha}} \ sech^{2}\Bigg[{\sqrt{{\frac{(u-c_{S})}{4B}}}}(x-ut)\Bigg]
\label{exactumKdV}
\end{equation}
where $u$ is an arbitrary supersonic velocity.

A soliton is a localized pulse which propagates without change in shape. On the other hand, 
the solutions of (\ref{bwqcdxitauXt}) and (\ref{bwqcdTfincomrefens}) will break, i.e., they 
will aquire an oscillating behavior and will be spread out, loosing localization. Whether or 
not a given physical system will support soliton propagation depends ultimately on its 
equation of state (in our case, on the function $\varepsilon= \varepsilon(\rho_B)$ or 
 $\varepsilon= \varepsilon(p)$). If the EOS takes into account the inhomogeneities in the 
system, the energy density will, in general, be a function of gradients and/or Laplacians.  
When used as input in   hydrodynamical equations, these higher order derivatives will, 
after some algebra, lead to the KdV equation.  In a hadronic phase, where the degrees of 
freedom are baryons and mesons, 
we have shown \cite{fn1,fn2,fn3,fn4} that the hydrodynamical equations will indeed 
give origin to the KdV equation.  In the present case, for this simple model of the quark 
gluon plasma this was not the case and we could only obtain the breaking wave equation. 

\subsection{Zero temperature}

Although the main focus of this work are the perturbations in a hot QGP formed in heavy ion 
collisions, for completeness, we discuss in this subsection the zero temperature case, which 
might be relevant for astrophysics. 

We will present numerical solutions of  (\ref{bwqcdxitauXt}) with the following initial  
condition, inspired by  (\ref{exactumKdV})
\begin{equation}
\hat\rho_{1}(x,t_{0})= A \ sech^{2}\bigg[\frac{x}{B}\bigg]
\label{exactumKdVTZEROlv}
\end{equation}
where $A$ and $B$ represent the amplitude and width  (of the initial baryon 
density pulse)  respectively. 
In Fig. 2 we show the numerical solution of (\ref{bwqcdxitauXt}) for $A=0.075$ 
and $B=1$ fm for different times. We can  observe the evolution of the initial 
gaussian-like pulse, the formation of a ``wall'' on the right side.  
\begin{figure}[h]
\begin{center}
\epsfig{file=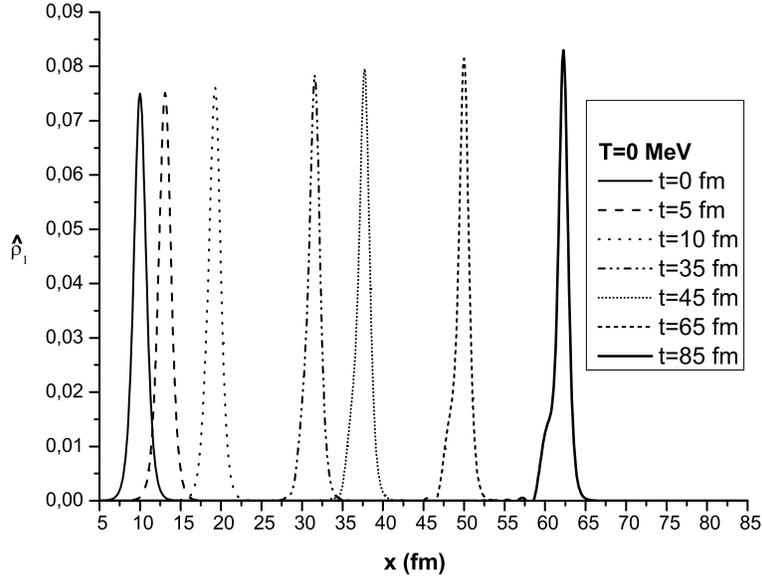,height=90mm}
\caption{Time evolution of a baryon density pulse at zero temperature.}
\end{center}
\label{fig2}
\end{figure}
Fig. 3  shows  the  numerical solution of (\ref{bwqcdxitauXt})  
for $A=0.35$  and $B=1$ fm. The time evolution of the pulse is similar to the 
one found in Fig. 2 but the ``wall'' formation and dispersion occurs much earlier. 
\begin{figure}[h]
\begin{center}
\epsfig{file=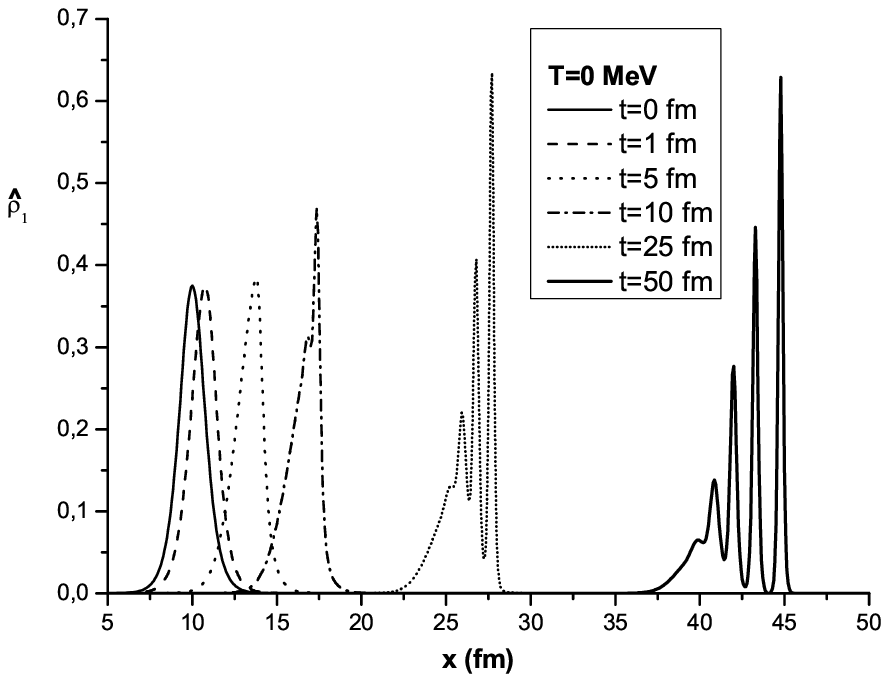,height=90mm}
\caption{The same as Fig. 2 for a larger amplitude.}
\end{center}
\label{fig3}
\end{figure}
In  Fig. 4  we present another solution of (\ref{bwqcdxitauXt}) 
for  $A=0.075$ and  $B=0.5$ fm.
\begin{figure}[h]
\begin{center}
\epsfig{file=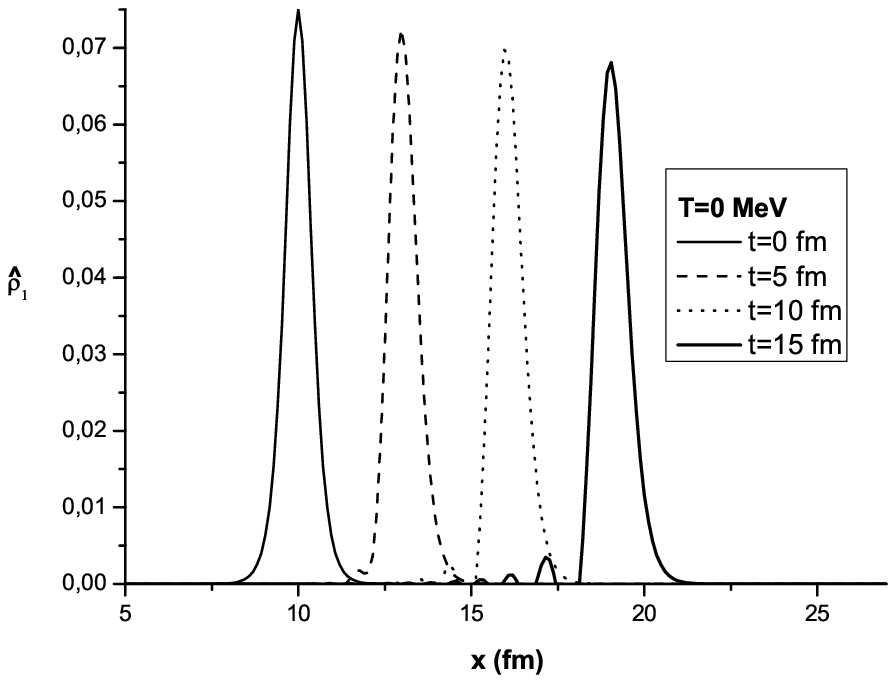,height=90mm}
\caption{The same as Fig. 2 for a smaller width.}
\end{center}
\label{fig4}
\end{figure}
We can see that the initial pulse starts to develop small secondary peaks, which are 
called ``radiation'' in the literature.  Further time evolution would increase the 
strenght of these peaks until the complete loss of localization. 

From these figures we learn how the solution depends on the initial amplitude and  
width: it lives longer as a compact pulse for smaller amplitudes and larger widths. 
Changes in one quantity may compensate the changes in the other, creating a very stable 
moving object. In fact,  the most striking conclusion to be drawn here is that for a 
wide variety of choices in the initial conditions the solution remains stable and 
localized for distances much larger than the nuclear size.


\subsection{Finite temperature}

We now turn to the study of the solutions of (\ref{bwqcdTfincomrefens}) for initial 
conditions given by (\ref{exactumKdVTZEROlv}) (replacing $\hat\rho_{1}$  by  
$\hat\varepsilon_{1}$). Now, beside the amplitude and width, the solution will depend also 
on the temperature.  When $T_{0}=T_{c}$ (\ref{bwqcdTfincomrefens})  reduces to:
\begin{equation}
{\frac{\partial\hat\varepsilon_{1}}{\partial t}}+c_{S}{\frac{\partial \hat\varepsilon_{1}}
{\partial x}}+
{\frac{2}{3}}{c_{S}}
\hat\varepsilon_{1}{\frac{\partial \hat\varepsilon_{1}}{\partial x}}=0 
\label{bwqcdTfincomrefenstzeroigualtece}
\end{equation}
which, changing the function from  $\hat\varepsilon_{1}$ to $\hat\rho_{1}$ is  
equal to (\ref{bwqcdxitauXt}). The previous conclusions are then extended to the 
present case. When  $T_{0}>>T_{c}$   (\ref{bwqcdTfincomrefens})  reduces to: 
\begin{equation}
{\frac{\partial\hat\varepsilon_{1}}{\partial t}}+c_{S}{\frac{\partial 
\hat\varepsilon_{1}}{\partial x}}+
{\frac{c_{S}}{2}}\hat\varepsilon_{1}{\frac{\partial \hat\varepsilon_{1}}{\partial x}}=0 
\label{bwqcdTfinaproxcomrefensdenovolim}
\end{equation}
Observing these two formulas we can see that, since $c_s = 1/3$ is fixed, the only change 
in the differential equation with temperature happens in the numerical coefficient of the 
last term which goes from $0.5$ to $0.66$. Therefore our results depend very weakly on the 
temperature. A stronger dependence on $T$ would appear if $c_s$ was allowed to change with 
temperature. This would correspond to having a different and more complicated equation of 
state for the quark gluon plasma.

In Fig. 5 we show the solution of  (\ref{bwqcdTfincomrefens}) with the initial condition 
given by  (\ref{exactumKdVTZEROlv}) with $A=0.01$,  $B=1$ fm and $T=300$ MeV. 
\begin{figure}[h]
\begin{center}
\epsfig{file=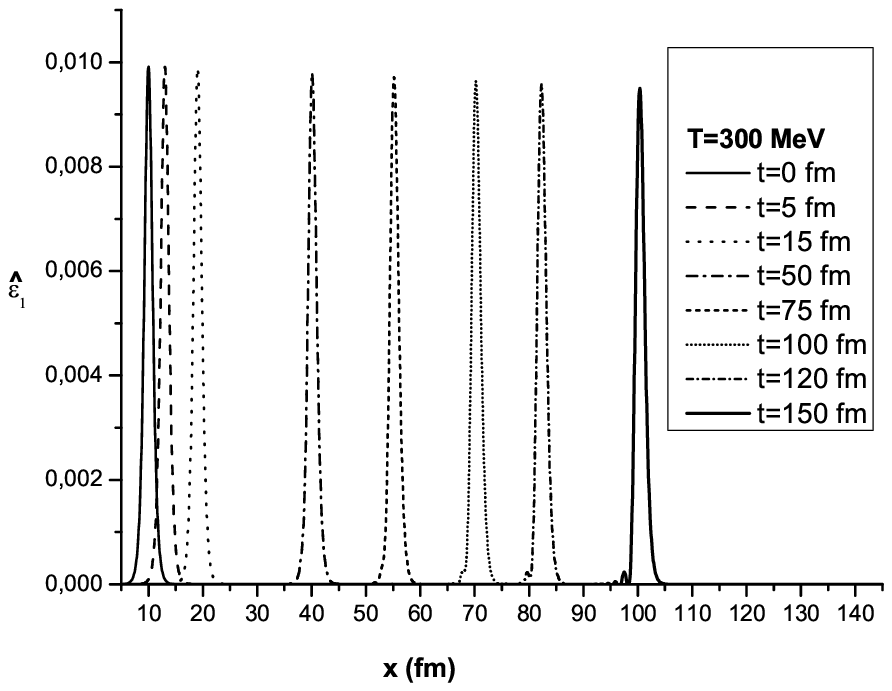,height=90mm}
\caption{Time evolution of an energy density pulse at $T=300$ MeV.}
\label{fig5}
\end{center}
\end{figure}
Fig. 6  shows the same as Fig. 5 but with $A=0.1$ and $B=1$ fm. As in the zero 
temperature case, we observe  that increasing the initial amplitude the breaking process 
develops earlier. 
\begin{figure}[h]
\begin{center}
\epsfig{file=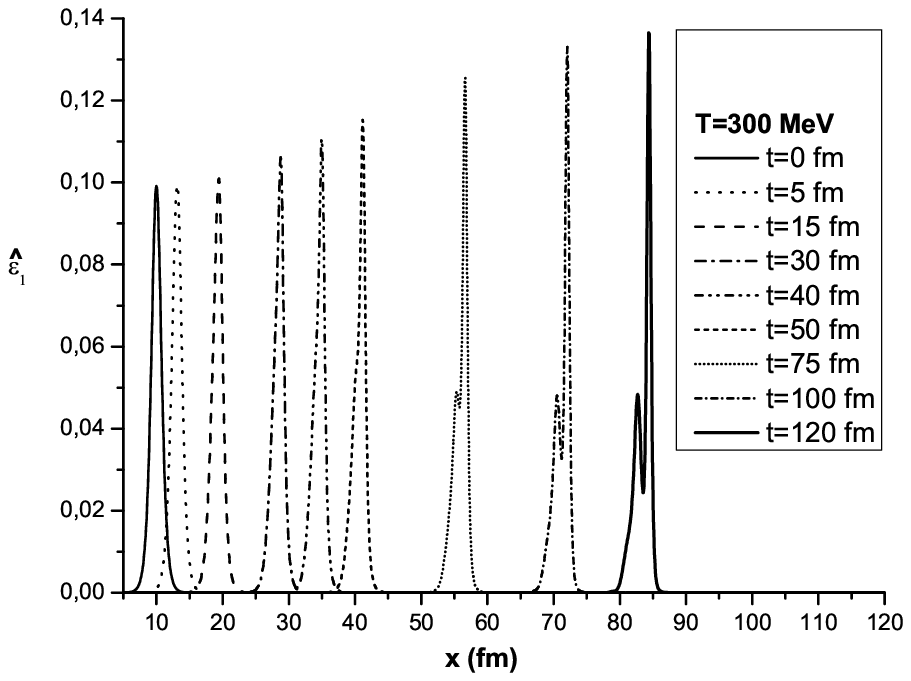,height=90mm}
\caption{The same as Fig. 5 for a larger amplitude.}
\label{fig6}
\end{center}
\end{figure}
In Fig. 7 we show the same as Fig. 5 but with $A=0.01$ and $B=0.2$ fm.
Figs. 8 and 9  show the time evolution of a pulse with the same initial 
amplitude ($A=0.5$) and width ($B=1$ fm) but different temperatures. Even though 
one temperature is $T=150$ MeV (Fig. 8) and the other is $T=300$ MeV (Fig. 9) we 
can hardly notice any difference.

\begin{figure}[h]
\begin{center}
\epsfig{file=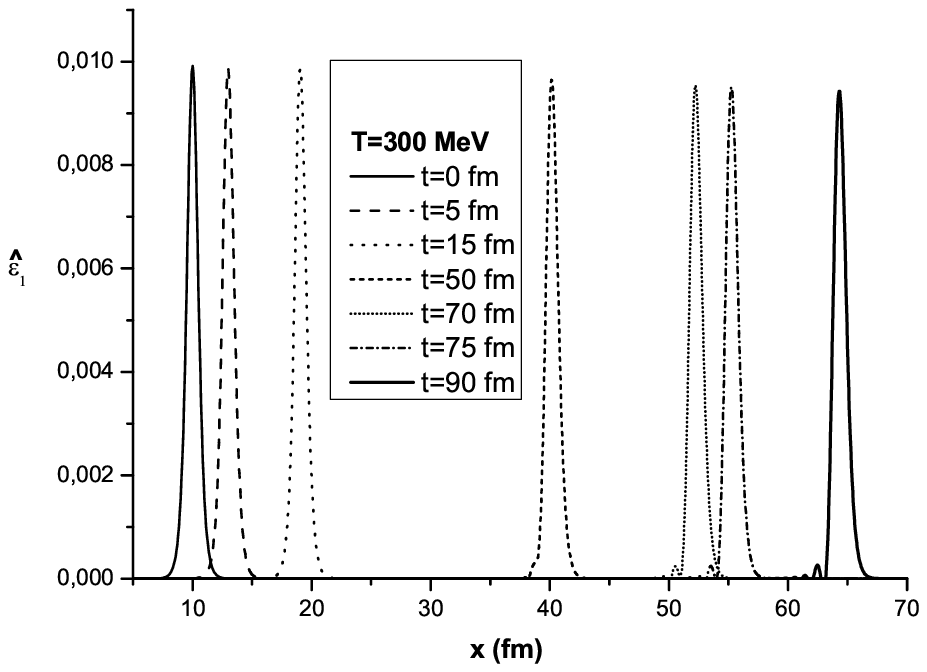,height=90mm}
\caption{The same as Fig. 5 for  a smaller width.}
\label{fig7}
\end{center}
\end{figure}

\begin{figure}[h]
\begin{center}
\epsfig{file=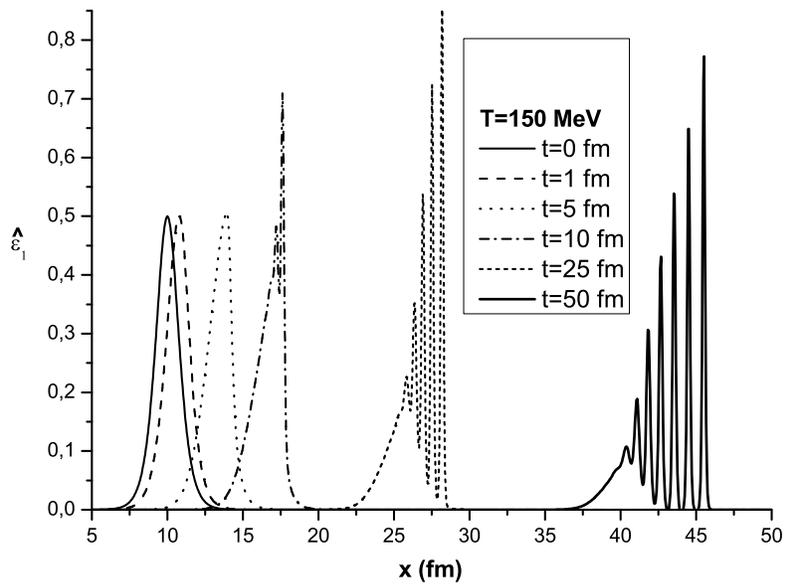,height=90mm}
\caption{Evolution of the energy density pulse at $T=150$ MeV.}
\end{center}
\label{fig8}
\end{figure}

\begin{figure}[h]
\begin{center}
\epsfig{file=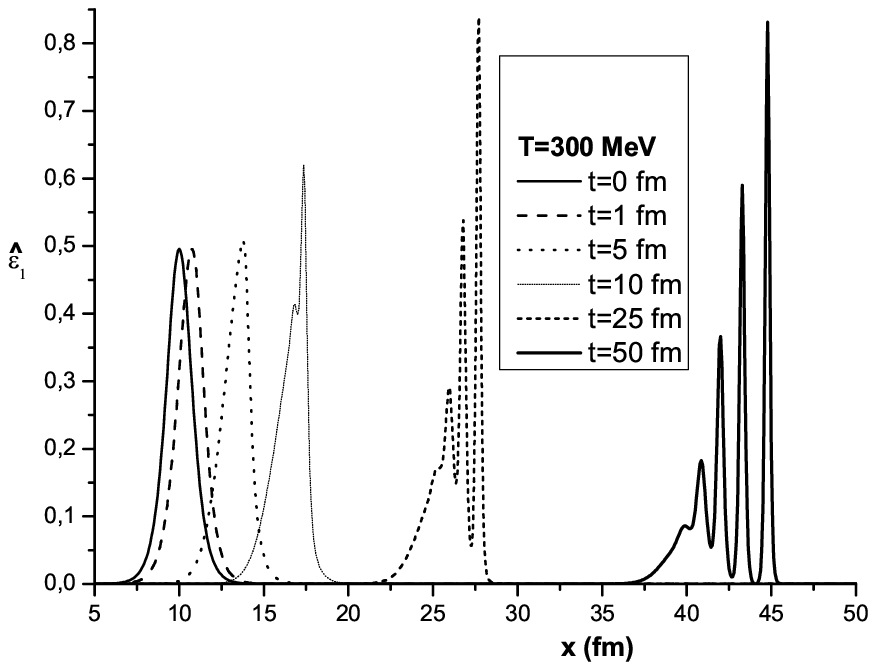,height=90mm}
\caption{The same as Fig. 8 for $T=300$ MeV.}
\label{fig9}
\end{center}
\end{figure}

\section{Conclusions}

We have proposed an alternative explanation  for the observed broadening of the 
away-side peak. It is based on the hydrodynamical treatment of energy perturbations. 
In contrast to other approaches we went beyond linearization of the fundamental equations
and did not neglect the non-linear terms.  We used a simple equation of 
state for the QGP and expanded the hydrodynamic equations around equilibrium 
configurations. The resulting differential equations describe the propagation of 
perturbations in the energy density. We solved them numerically and found that 
localized perturbations can propagate for long distances in the plasma. Under 
certain conditions our solutions mimick the propagation of  Korteweg - de Vries 
solitons. However, as said before, from this finding to a realistic calculation and a 
serious  attempt to describe the data there is still a long way. The main result found
in this work, namely, the persistence of soliton-like configurations, is very promising 
and encourages us to extend our formalism to two spatial dimensions. This project  is 
 in progress.
\begin{acknowledgments}
We are deeply grateful to S. Raha for useful discussions.
This work was  partially financed by the Brazilian funding
agencies CAPES, CNPq and FAPESP. 
\end{acknowledgments}

\end{document}